%% file: erice99.tex
\documentclass[12pt]{article}
\input{psfig}
 \usepackage{pictex}
 \usepackage{epsfig}
 \usepackage{pstricks}
 \usepackage{pst-coil}

\def\b{\beta}

\def\d{\delta}
\def\e{\epsilon}
\def\f{\phi}
\def\g{\gamma}

\def\l{\lambda}
\def\m{\mu}
\def\n{\nu}

\def\p{\pi}

\def\r{\rho}

\def\t{\tau}

\def\x{\xi}

\def\D{\Delta}

\def\G{\Gamma}

\def\cl{{\cal L}}
\def\cm{{\cal M}}

\def\co{{\cal O}}

\def\bo{{\raise.15ex\hbox{\large$\Box$}}}               
\def\pr{\prod}                                          
\def\face{{\raise.2ex\hbox{$\displaystyle \bigodot$}\mskip-2.2mu \llap {$\ddot
        \smile$}}}                                      
\def\beq{\begin{equation}}
\def\eeq{\end{equation}}

\def\bea{\begin{eqnarray}}
\def\eea{\end{eqnarray}}
\def\NO{\nonumber}

\def\leftrightarrowfill{$\mathsurround=0pt \mathord\leftarrow \mkern-6mu
        \cleaders\hbox{$\mkern-2mu \mathord- \mkern-2mu$}\hfill
        \mkern-6mu \mathord\rightarrow$}       
\def\dvec#1{\vbox{\ialign{##\crcr
        \leftrightarrowfill\crcr\noalign{\kern-1pt\nointerlineskip}
        $\hfil\displaystyle{#1}\hfil$\crcr}}}           

\def\pl#1#2#3{Phys.~Lett.~{\bf B {#1}} (19{#2}) #3}
\def\np#1#2#3{Nucl.~Phys.~{\bf B {#1}} (19{#2}) #3}
\def\prl#1#2#3{Phys.~Rev.~Lett.~{\bf #1} (19{#2}) #3}
\def\pr#1#2#3{Phys.~Rev.~{\bf D {#1}} (19{#2}) #3}

\def\prep#1#2#3{Phys.~Rep.~{\bf {#1}C} (19{#2}) #3}

\catcode`@=11
\def\@citex[#1]#2{\if@filesw\immediate\write\@auxout{\string\citation{#2}}\fi
  \def\@citea{}\@cite{\@for\@citeb:=#2\do
    {\@citea\def\@citea{,\penalty\@m}\@ifundefined
      {b@\@citeb}{{\bf ?}\@warning
       {Citation `\@citeb' on page \thepage \space undefined}}%
\hbox{\csname b@\@citeb\endcsname}}}{#1}}

\def\citer{\@ifnextchar [{\@tempswatrue\@citexr}{\@tempswafalse\@citexr[]}}
\def\@citexr[#1]#2{\scriptsize 
  \if@filesw\immediate\write\@auxout{\string\citation{#2}}\fi
  \def\@citea{}\@cite{\@for\@citeb:=#2\do
    {\@citea\def\@citea{-\penalty\@m}\@ifundefined
       {b@\@citeb}{{\bf ?}\@warning
       {Citation `\@citeb' on page \thepage \space undefined}}%
\hbox{\csname b@\@citeb\endcsname}}}{#1}\normalsize}
\catcode`@=12

\catcode`\@=11
\long\def\@makefntext#1{ 
\protect\noindent \hbox to 3.2pt {\hskip-.9pt  
$^{{\ninerm\@thefnmark}}$\hfil}#1\hfill} 

 \def\@makefnmark{\hbox to 0pt{$^{\@thefnmark}$\hss}}  
        
\def\ps@myheadings{\let\@mkboth\@gobbletwo
\def\@oddhead{\hbox{} 
\rightmark\hfil\ninerm\thepage}   
\def\@oddfoot{}\def\@evenhead{\ninerm\thepage\hfil 
\leftmark\hbox{}}\def\@evenfoot{}
\def\sectionmark##1{}\def\subsectionmark##1{}}

\begin{document}

\date{}
\title{
{\normalsize\rm DESY 00-004}\hfill{\mbox{}}\\
{\normalsize\rm January 2000}\hfill{\mbox{}}\vspace*{2cm}\\
{\bf ELEMENTS OF BARYOGENESIS\footnote{Presented at the International School 
of Astrophysics D. Chalonge, Erice, December 1999}}
}
\author{W.~Buchm\"uller and S.~Fredenhagen  \\
\vspace{3.0\baselineskip}                                               
{\normalsize\it Deutsches Elektronen-Synchrotron DESY, 22603 Hamburg, Germany}
\vspace*{1cm}                     
}        

\maketitle

\thispagestyle{empty}

\begin{abstract}
\noindent
Basic ingredients of the theory of baryogenesis are reviewed with emphasis on 
out-of-equilibrium decays of heavy particles. The present use of kinetic
theory is explained and some attempts to go beyond the classical Boltzmann
equations are discussed.
\end{abstract}

\newpage

\section{Scenarios for baryogenesis}

One of the main successes of the standard early-universe cosmology is the
prediction of the abundances of the light elements, D, $^3$He, $^4$He and 
$^7$Li. Agreement between theory and observation is obtained for
a certain range of the parameter $\eta$, the ratio of baryon density and
photon density \cite{pdb98},
\beq
\eta = {n_B\over n_\g} = (1.5 - 6.3)\times 10^{-10}\;,
\eeq
where the present number density of photons is $n_\g \sim 400/{\rm cm}^3$. 
Since no significant amount of antimatter is observed in the universe, 
the baryon density yields directly the cosmological baryon asymmetry, 
$n_B \simeq n_B - n_{\bar{B}}$.

A matter-antimatter asymmetry can be dynamically generated in an expanding
universe if the particle interactions and the cosmological evolution satisfy 
Sakharov's conditions \cite{sa67}, i.e.
\begin{itemize}
\item baryon number violation
\item $C$ and $C\!P$ violation
\item deviation from thermal equilibrium .
\end{itemize}
Although the baryon asymmetry is just a single number, it provides an
important relationship between the standard model of cosmology, i.e. the
expanding universe with Robertson-Walker metric, and the standard model
of particle physics as well as its extensions.

At present there are a number of viable scenarios for baryogenesis. They
can be classified according to the different ways in which Sakharov's 
conditions are realized. Already in the standard model $C$ and $C\!P$ are
not conserved. Also baryon number ($B$) and lepton number ($L$) are
violated by instanton processes \cite{tho76}. In grand unified theories
$B$ and $L$ are broken by the interactions of gauge bosons and leptoquarks.
This is the basis of the classical GUT baryogenesis \cite{yo78}.
Analogously, the $L$ violating decays of heavy Majorana neutrinos lead to
leptogenesis \cite{fy86}. In supersymmetric theories the existence of 
approximately flat directions in the scalar potential leads to new 
possibilities. Coherent oscillations of scalar fields may then generate
large asymmetries \cite{ad85}.

The crucial departure from thermal equilibrium can also be realized in several
ways. One possibility is a sufficiently strong first-order electroweak phase 
transition \cite{rs96}. In this case $C\!P$ violating interactions of the
standard model or its supersymmetric extension could in principle generate the
observed baryon asymmetry. However, due to the rather large lower bound on the
Higgs boson mass of about 105~GeV, which is imposed by the LEP experiments,
this interesting possibility is now restricted to a very small range of 
parameters in the supersymmetric standard model. In the case of the 
Affleck-Dine scenario the baryon asymmetry is generated at the end of an 
inflationary period as a coherent effect of scalar fields which leads to an
asymmetry between quarks and antiquarks after reheating \cite{drt96}.
For the classical GUT baryogenesis and for leptogenesis the departure from
thermal equilibrium is due to the deviation of the number density of the
decaying heavy particles from the equilibrium number density.
How strong this deviation from thermal equilibrium is depends on the lifetime
of the decaying heavy particles and the cosmological evolution. Further
scenarios for baryogenesis are described in \cite{do92}.

The theory of baryogenesis involves non-perturbative aspects of quantum
field theory and also non-equilibrium statistical field theory, in particular
the theory of phase transitions and kinetic theory. 
\begin{figure}[h]
\begin{center}
\scaleboxto(7,7) {\parbox[c]{9cm}{\input{Fig01.tex}}}
\end{center}
\caption{\it One of the 12-fermion processes which are in thermal 
equilibrium in the high-temperature phase of the standard model.
\label{fig_sphal}}
\end{figure}
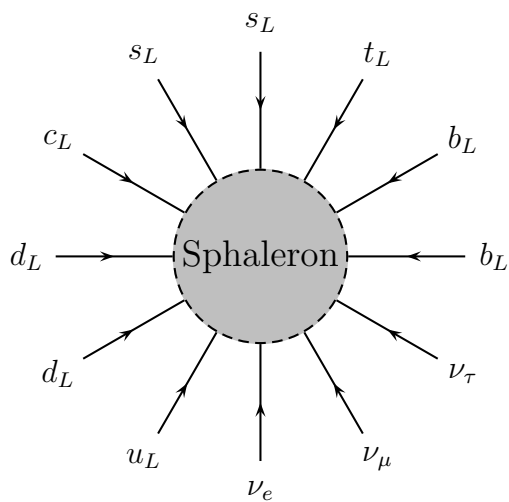
A crucial ingredient is the connection between baryon number and lepton 
number in the high-temperature, symmetric phase of
the standard model. Due to the chiral nature of the weak interactions $B$ and
$L$ are not conserved. At zero temperature this has no observable 
effect due to the smallness of the weak coupling. However, as the temperature 
approaches the critical temperature $T_{EW}$ of the electroweak transition, 
$B$ and $L$ violating processes come into thermal equilibrium \cite{krs85}. 

The rate of these processes is
related to the free energy of sphaleron-type field configurations which carry
topological charge. In the standard model they lead to an effective
interaction of all left-handed fermions \cite{tho76} 
(cf. fig.~\ref{fig_sphal}), 
\beq\label{obl}
O_{B+L} = \prod_i \left(q_{Li} q_{Li} q_{Li} l_{Li}\right)\; ,
\eeq
which violates baryon and lepton number by three units, 
\beq 
    \D B = \D L = 3\;. \label{sphal1}
\eeq
The evaluation of the sphaleron rate in the symmetric high temperature phase 
is a challenging problem \cite{asy97}. Although a complete theoretical 
understanding has not yet been achieved, it is generally believed that $B$ 
and $L$ violating processes are in thermal equilibrium for temperatures in 
the range
\beq 
T_{EW} \sim 100\ \mbox{GeV} < T < T_{SPH} \sim 10^{12}\ \mbox{GeV}\;.
\eeq

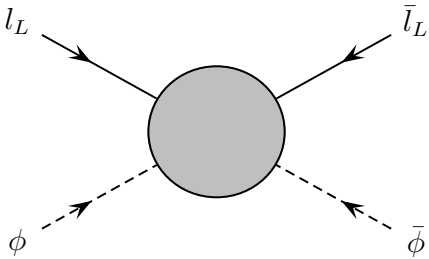
\begin{figure}[h]
\begin{center}
\scaleboxto(7,3.5){
\parbox[c]{9cm}{\input{Eff_Interaction.tex}}}
\end{center}
\caption{\it Effective lepton number violating interaction.\label{fig_lept}}
\end{figure}

Sphaleron processes have a profound effect on the generation of the
cosmological baryon asymmetry, in particular in connection with lepton
number violating interactions between lepton and Higgs fields,
\beq\label{dl2}
\cl_{\Delta L=2} ={1\over 2} f_{ij}\ l^T_{Li}\phi\ C\ l_{Lj}\phi 
                  +\mbox{ h.c.}\;.\label{intl2}
\eeq
Such an interaction arises in particular from the exchange of heavy Majorana
neutrinos (cf.~fig.~\ref{fig_lept}). In the Higgs phase of the standard
model, where the Higgs field acquires a vacuum expectation value, it gives
rise to Majorana masses of the light neutrinos $\n_e$, $\n_\m$ and $\n_\t$.   

Eq.~(\ref{sphal1}) suggests that any
$B+L$ asymmetry generated before the electroweak phase transition,
i.e., at temperatures $T>T_{EW}$, will be washed out. However, since
only left-handed fields couple to sphalerons, a non-zero value of
$B+L$ can persist in the high-temperature, symmetric phase if there
exists a non-vanishing $B-L$ asymmetry. An analysis of the chemical potentials
of all particle species in the high-temperature phase yields a
relation between the baryon asymmetry $Y_B = (n_B-n_{\bar{B}})/s$, where $s$
is the entropy density, and the corresponding $B-L$ asymmetry $Y_{B-L}$, 
respectively~\cite{ht90},
\beq\label{basic}
Y_B\ =\ C\ Y_{B-L}\ =\ {C\over C-1}\ Y_L\;.
\eeq
The number $C$ depends on the other processes which are in thermal equilibrium~\cite{bp99}. 
If these are all standard model interactions one has 
$C=(8N+4)/(22N+13)$ for $N$ generations. If instead of the Yukawa interactions
of the right-handed electron the $\D L=2$ interactions (\ref{dl2}) are in
equilibrium one finds $C=-2N/(2N+3)$.

The interplay between the sphaleron processes (fig.~\ref{fig_sphal}) 
and the lepton number changing processes (fig.~\ref{fig_lept}) leads to an
intriguing relation between neutrino properties and the cosmological baryon 
asymmetry. The decay of heavy Majorana neutrinos can quantitatively account
for the observed asymmetry.

\section{Heavy particle decays in a thermal bath}
Let us now consider the simplest possibility for a departure from thermal
equilibrium, the decay of heavy, weakly interacting particles in a thermal
bath. To be specific, we choose the heavy particle to be a
Majorana neutrino $N=N^c$ which can decay into a lepton Higgs  
pair $l \phi$ and also into the $C\!P$ conjugate state $\bar{l} \bar{\phi}$\,
\beq
N \rightarrow l\;\phi\;, \quad N \rightarrow \bar{l}\;\bar{\phi}\;.
\eeq
In the case of $C\!P$ violating couplings a lepton asymmetry can be generated in
the decays of the heavy neutrinos $N$ which is then partially transformed
into a baryon asymmetry \cite{fy86} by sphaleron processes \cite{krs85}. 
Compared to other scenarios of baryogenesis this leptogenesis mechanism
has the advantage that, at least in principle, the resulting baryon asymmetry
is entirely determined by neutrino properties. 

\begin{figure}[ht]
\begin{center}
\scaleboxto(5.4,0){\parbox[c]{9cm}{\input{decay1.tex}}}
\scaleboxto(5.4,0){\parbox[c]{9cm}{\input{decay2.tex}}}
\scaleboxto(5.4,0){\parbox[c]{9cm}{\input{decay3.tex}}}
\scaleboxto(5.4,0){\parbox[c]{9cm}{\input{decay4.tex}}}
\end{center}
\caption{\it $\D L=1$ processes: decays and inverse decays of a heavy Majorana
neutrino.\label{decinv}}
\end{figure}
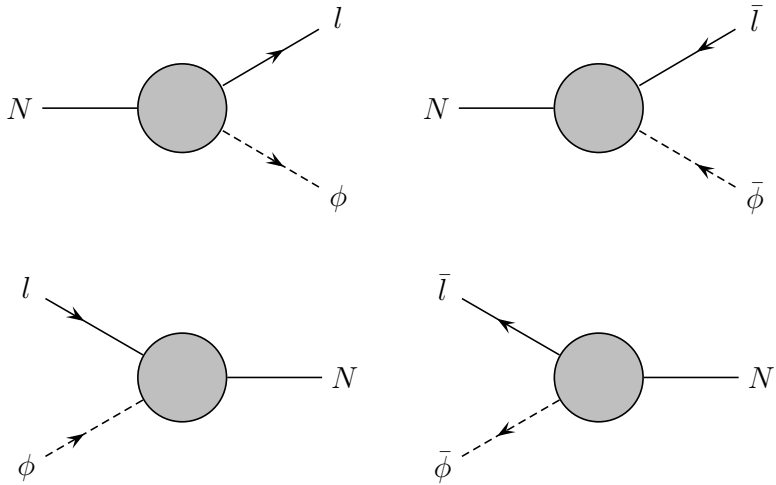
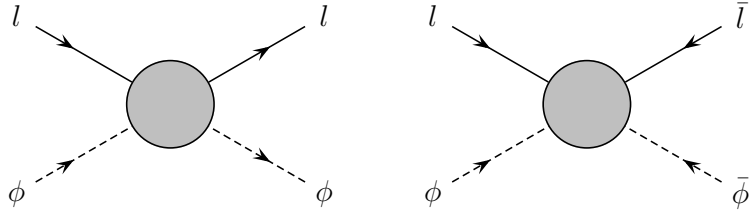
\begin{figure}[ht]
\begin{center}
\scaleboxto(5.4,0){
\parbox[c]{9cm}{\input{twotwo.tex}}}
\scaleboxto(5.4,0){
\parbox[c]{9cm}{\input{twoanti.tex}}}
\end{center}
\caption{\it $\D L=0$ and $\D L=2$ lepton Higgs processes.\label{lephig}}
\end{figure}

The generation of a baryon asymmetry is an out-of-equilibrium process 
which is generally treated by means of Boltzmann equations. A detailed
discussion of the basic ideas and some of the subtleties has been given
in \cite{kw80}. The main processes in the thermal bath are the decays
and the inverse decays of the heavy neutrinos (cf.~fig.~\ref{decinv}), and
the lepton number conserving ($\D L=0$) and violating ($\D L=2$) processes
(cf.~fig.~\ref{lephig}). In addition there are other processes, in particular
those involving the t-quark, which are also important in a quantitative
analysis \cite{lu92,pl97}. A lepton asymmetry can be dynamically generated
in an expanding universe if the partial decay widths of the heavy neutrino
do not respect $C\!P$ symmetry,
\beq\label{gcp} 
\G(N\rightarrow l\phi) = {1\over 2}(1+\e)\G\;,\quad
\G(N\rightarrow \bar{l}\bar{\phi})={1\over 2}(1-\e)\G\;,
\eeq
where $\G$ is the total decay width and the parameter $\e \ll 1$ measures the 
amount of $C\!P$ violation.

The Boltzmann equations for the number densities of heavy neutrinos ($n_N$),
leptons ($n_l$) and antileptons ($n_{\bar{l}}$) corresponding to the 
processes in figs.~\ref{decinv} and \ref{lephig} are given by
\bea
{d n_N\over dt} + 3 H n_N &=& -\g(N \rightarrow l\phi) 
                              + \g(l\phi \rightarrow N) \NO\\
&&  -\g(N \rightarrow \bar{l}\bar{\phi})
    +\g(\bar{l}\bar{\phi} \rightarrow N)\;,\label{nN}\\
{d n_l\over dt} + 3 H n_l &=& \g(N \rightarrow l\phi) 
                              - \g(l\phi \rightarrow N) \NO\\
&& + \g(\bar{l}\bar{\phi}\rightarrow l\phi)
   - \g(l\phi \rightarrow \bar{l}\bar{\phi})\;,\label{nl}\\
{d n_{\bar{l}}\over dt} + 3 H n_{\bar{l}} 
&=& \g(N \rightarrow \bar{l}\bar{\phi})
    -\g(\bar{l}\bar{\phi}\rightarrow N) \NO\\
&&  +\g(l\phi \rightarrow \bar{l}\bar{\phi})
    - \g(\bar{l}\bar{\phi}\rightarrow l\phi)\;,\label{nbarl}
\eea
with the reaction rates
\bea
\g(N\rightarrow l\phi) &=& \int d\Phi_{123} 
      f_N(p_1) |\cm(N\rightarrow l\phi)|^2\;, \ldots \label{nlphi} \\
\g(l\phi\rightarrow \bar{l}\bar{\phi}) &=& \int d\Phi_{1234}
 f_l(p_1)f_{\phi}(p_2) 
 |\cm'(l\phi\rightarrow \bar{l}\bar{\phi})|^2\;,\ldots\label{llphi}
\eea
Here $H$ is the Hubble parameter, $d\Phi_{1...n}$ denotes the phase space 
integration over particles in initial and final states,
\beq
d\Phi_{1...n}={d^3p_1\over (2\pi)^3 2E_1}\ldots {d^3p_n\over (2\pi)^3 2E_n}
(2\pi)^4 \d^4(p_1+\ldots -p_n)\;,
\eeq
and 
\beq
f_i(p)=\exp{(-\b E_i(p))}\;, \quad 
n_i(p) = g_i \int {d^3p\over (2\pi)^3} f_i(p)\;,
\eeq
are Boltzmann distribution and number density of particle $i=N,l,\phi$ at
temperature $T=1/\b$, respectively. $\cm$ and $\cm'$ denote the scattering 
matrix elements of the indicated processes at zero temperature; the prime
indicates that for the $2\rightarrow 2$ processes the contribution of the 
intermediate resonance state has been subtracted. For simplicity we have used 
in eqs.~(\ref{nlphi}) and
(\ref{llphi}) Boltzmann distributions rather than Bose-Einstein and
Fermi-Dirac distributions, and we have also neglected the distribution
functions in the final state which is a good approximation for small
number densities. Subtracting (\ref{nbarl}) from (\ref{nl}) yields the
Boltzmann equation for the asymmetry $n_l - n_{\overline{l}}$. 

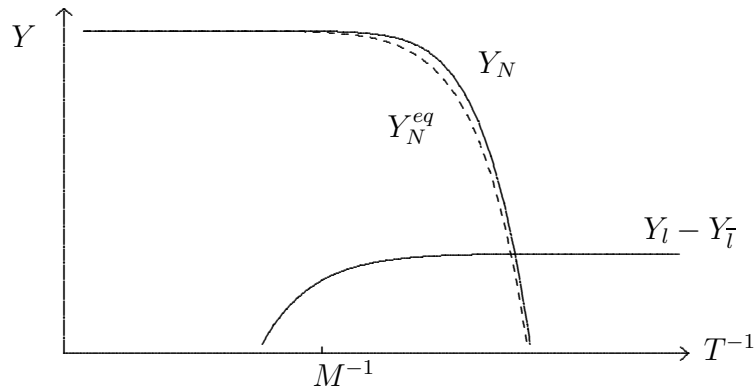
\begin{figure}
\begin{center}
\input{outofequ}
\end{center}
\caption{\label{fig:outofequ}\it Time evolution of the number density to 
entropy density ratio. At $T\sim M$ the system gets out of
equilibrium and an asymmetry is produced.}
\end{figure}

A typical solution of the Boltzmann equations (\ref{nN})~-~(\ref{nbarl}) is
shown in fig.~\ref{fig:outofequ}. Here the ratios of number densities and
entropy density,
\beq
Y_X = {n_X\over s}\;,
\eeq
are plotted, which remain constant in an expanding universe in thermal 
equilibrium. A heavy neutrino, which is weakly coupled 
to the thermal bath, falls out of thermal equilibrium at temperatures 
$T \sim M$ since its decay is too slow to follow the rapidly decreasing 
equilibrium distribution $f_N \sim \exp(-\b M)$. This leads to an excess of 
the number density, $n_N > n_N^{eq}$. $C\!P$ violating partial decay widths then yield a lepton
asymmetry which, by means of sphaleron processes, is partially transformed
into a baryon asymmetry.

The Boltzmann equations are classical equations for the time evolution of
number densities. The collision terms, however, are $S$-matrix elements which
involve quantum mechanical interferences of different amplitudes in a
crucial manner. Since these scattering matrix elements are evaluated at
zero temperature, one may worry to what extent the quantum mechanical
interferences are affected by interactions with the thermal bath. Another
subtlety is the separation of the $2\rightarrow 2$ matrix elements into
a resonance contribution and remainder \cite{kw80},
\beq\label{split}
|\cm(l\phi\rightarrow \bar{l}\bar{\phi})|^2 = 
|\cm'(l\phi\rightarrow \bar{l}\bar{\phi})|^2 + 
|\cm_{res}(l\phi\rightarrow \bar{l}\bar{\phi})|^2\;,
\eeq
where the resonance contribution has the form
\beq
\cm_{res}(l\phi\rightarrow \bar{l}\bar{\phi}) \propto
\cm(l\phi\rightarrow N)\cm(N\rightarrow \bar{l}\bar{\phi})^* =
|\cm(l\phi\rightarrow N)|^2\;.
\eeq 
The entire effect of baryon number generation crucially depends on this
separation. The particles which participate in the $2\rightarrow 2$ processes
are massless, hence their distribution functions always coincide with the
equilibrium distribution. Only the resonances, treated as on-shell particles,
fall out of thermal equilibrium and can then generate an asymmetry in their
decays. General theoretical arguments require cancellations between these
two types of contributions which we illustrate in the following with two
examples.\\
\smallskip

\noindent\textbf{Cancellations in thermal equilibrium}\\ \nopagebreak 

If all processes, including those which violate baryon number, are in
thermal equilibrium the baryon asymmetry vanishes. This is a direct
consequence of the $C\!PT$ invariance of the theory,
\bea
\langle B \rangle &=& \mbox{Tr}(\r B) 
=\mbox{Tr}\left((C\!PT)(C\!PT)^{-1}\exp{(-\b H)}B\right) \NO\\
&=&\mbox{Tr}\left(\exp{(-\b H)}(C\!PT)^{-1}B(C\!PT)\right) = -\mbox{Tr}(\r B)=0\;.
\eea
Hence, no asymmetry can be generated in equilibrium, and the
transition rate which determines the change of the asymmetry has to vanish,
\beq
{d(n_l-n_{\bar{l}})\over dt} + 3 H(n_l-n_{\bar{l}}) = \D\g^{eq} = 0\;,
\eeq
where the superscript $eq$ denotes rates evaluated with equilibrium 
distributions.

From eqs.~(\ref{gcp}), (\ref{nl}) and (\ref{nbarl}) one obtains for the 
resonance contribution, i.e. decay and inverse decay,
\beq
\D\g^{eq}_{res}=-2\e\g^{eq}(N\rightarrow l\phi)\;.
\eeq
This means in particular that the asymmetry generated in the decay is not
compensated by the effect of inverse decays. On the contrary, both processes
contribute the same amount.

The rate $\D\g^{eq}_{res}$ has to be compensated by the contribution from
$2\rightarrow 2$ processes which is given by
\beq
\D\g^{eq}_{2\rightarrow 2} = 2\int d\Phi_{1234}f_l^{eq}(p_1)f_{\phi}^{eq}(p_2)
\left(|\cm'(l\phi\rightarrow \overline{l}\overline{\phi})|^2 -
      |\cm'(\overline{l}\overline{\phi}\rightarrow l\phi)|^2\right)\;.
\eeq 
For weakly coupled heavy neutrinos, i.e. $\G \propto \l^2 M$ with $\l^2\ll 1$,
this compensation can be easily shown using the unitarity of the $S$-matrix.

The sum over states in the unitarity relation,
\beq
\sum_X\left(|\cm(l\phi\rightarrow X)|^2-|\cm(X\rightarrow l\phi)|^2\right)=0\;,
\eeq
can be restricted to two-particle states to leading order in the case of weak 
coupling $\l$. This implies for the considered $2\rightarrow 2$ processes,
\beq\label{sum}
{\sum_{l,\phi,\bar{l'},\bar{\phi'}}\!\!\!}'
\left(|\cm(l\phi\rightarrow \bar{l'}\bar{\phi'})|^2 -
      |\cm(\bar{l'}\bar{\phi'}\rightarrow l\phi)|^2 \right) = 0\;,
\eeq
where the summation $\sum'$ includes momentum integrations under the 
constraint of fixed total momentum. From eqs.~(\ref{split}) and (\ref{sum})
one obtains        
\beq
\D\g^{eq}_{2\rightarrow 2} = 2\int d\Phi_{1234}f_l^{eq}(p_1)f_{\phi}^{eq}(p_2)
\left(-|\cm_{res}(l\phi\rightarrow \bar{l}\bar{\phi})|^2 +
      |\cm_{res}(\bar{l}\bar{\phi}\rightarrow l\phi)|^2\right)\;.
\eeq 
In the narrow width approximation, i.e. to leading order in $\l^2$, this
yields the wanted result,
\bea
\D\g^{eq}_{2\rightarrow 2} &=& 
2\int d\Phi_{1234}f_l^{eq}(p_1)f_{\phi}^{eq}(p_2)
\left(-|\cm(l\phi\rightarrow N)|^2 |\cm(N\rightarrow \bar{l}\bar{\phi})|^2
\right.\NO\\
&&\left. \hspace{2cm}
+|\cm(\bar{l}\bar{\phi}\rightarrow N)|^2|\cm(N\rightarrow l\phi)|^2 \right)
{\pi\over M \G}\d(s-M^2) \NO\\
&=&2 \e \g^{eq}(N\rightarrow l\phi) = - \D\g^{eq}_{res}\;.
\eea
This cancellation also illustrates that the Boltzmann equations treat 
resonances as on-shell real particles. Off-shell effects require a different
formalism which will be discussed in Sec.~3.\\
\smallskip

\noindent\textbf{Cancellations at zero temperature}\\

\begin{figure}[ht]
\input{vertex.tex}
\caption{\it One-loop vertex correction to lepton Higgs scattering. 
\label{vertex}}
\end{figure}
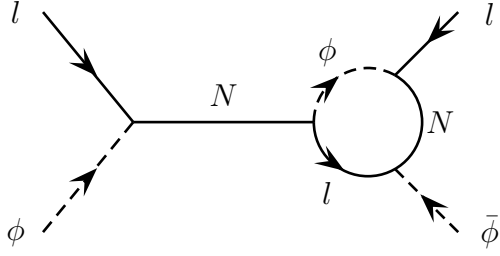

The lepton asymmetry which is obtained by solving the Boltzmann equations
depends crucially on the separation of the $2\rightarrow 2$ scattering
amplitudes into resonance contribution and remainder. How to perform this
separation appears obvious for a graph like fig.~\ref{vertex} which 
represents the interference between the tree level and a one-loop correction 
term for the vertex $\l Nl\phi$, together with a $s$-channel $N$-propagator. 

\begin{figure}[ht]
\input{selfen.tex}
    \caption{\it One-loop self-energy correction to lepton Higgs scattering. 
\label{selfen}}
 \end{figure}
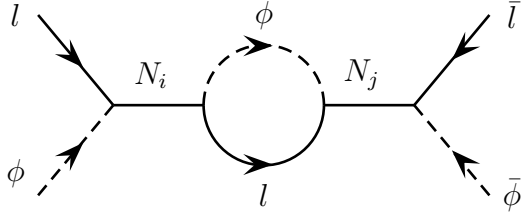

The identification of the resonance contribution shown in fig.~\ref{selfen}
is less obvious. One expects mixing effects in the case of several
heavy neutrinos which have been discussed in \cite{ls93,fps95,crv96}.
Treating the one-loop self-energy like the one-loop vertex correction
yields indeed a finite contribution to the $C\!P$ asymmetry \cite{fps95,crv96}
which is of the same order as the vertex contribution. However, one may
worry about the fact that the same procedure yields an infinite result for the
total and partial decay widths. The self-energy corrections have to be
resummed in order to determine the mass of the heavy neutrino, its partial 
decay widths and, in particular, the $C\!P$ asymmetry.
It is well known that the properties of unstable particles are defined by 
position and residue of the corresponding poles of the scattering 
matrix \cite{ve63}. These poles correspond to the poles of the full propagator.

It turns out that the resummed propagator does not contribute to the $C\!P$ 
asymmetry of $2\rightarrow 2$ processes at fixed external momenta \cite{bp98} 
(fig.~\ref{asprop}). 
\begin{figure}[ht]\begin{center}
\scaleboxto(14.4,0){\parbox{16cm}{\input{prop.tex}}}\end{center}
\caption{\it Propagator contribution to $C\!P$ asymmetry for fixed external 
momenta. \label{asprop}}
\end{figure}
Furthermore, even the total $C\!P$ asymmetry vanishes to leading order in 
$\l^2$ when integrated over phase space~\cite{rcv98} (fig.~\ref{asfull}). 
\begin{figure}[ht]
\input{full.tex}
\caption{\it Total $C\!P$ asymmetry to leading non-trivial order in the
coupling. \label{asfull}}
\end{figure}
This can be seen explicitly in ordinary perturbation theory \cite{rcv98} for 
center-of-mass energies below the resonance region, $s\ll M^2$, 
as well as in the resonance region, $s \sim M^2$, which can 
only be studied after resummation \cite{bp98}.

The cancellation of the various contributions to the $C\!P$ asymmetry follows 
directly from unitarity. In fact, this is the physical meaning of 
eq.~({\ref{sum}),
\[
{\sum_{l,\phi,\bar{l'},\bar{\phi'}}\!\!\!}'
\left(|\cm(l\phi\rightarrow \bar{l'}\bar{\phi'})|^2 -
      |\cm(\bar{l'}\bar{\phi'}\rightarrow l\phi)|^2 \right) = 0\;. 
\]
Away from resonance poles, where ordinary perturbation theory holds, the 
$C\!P$ asymmetry  vanishes to order $\l^6$. Corrections due to 
four-particle intermediate states are $\co (\l^8)$. In the resonance region
the $C\!P$ asymmetry vanishes to order $\l^2$ with corrections $\co (\l^4)$.

These cancellations between various contributions to the $C\!P$ asymmetry 
demonstrate the importance of identifying correctly the resonance 
contribution. This is complicated by the fact that different
chiral projections of the resummed propagator are diagonalized by
different unitary matrices \cite{bp98}. These matrices, together with the 
vertex corrections, determine the effective $Nl\phi$ vertex (fig.~\ref{fullv}),
where $N$ now corresponds to a pole of the full heavy neutrino propagator.
\begin{figure}[ht]
\input{fullv.tex}
\caption{\it Effective $Nl\phi$ vertex including mixing effects. \label{fullv}}
\end{figure}
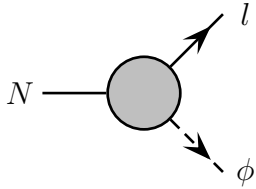

Given the effective $Nl\phi$ vertex it is straightforward to determine the 
$C\!P$ asymmetry in the decay of a heavy Majorana neutrino,
\beq
\e_i = {\G(N_i\rightarrow l\f) - \G(N_i\rightarrow \bar{l}\bar{\f})\over 
        \G(N_i\rightarrow l\f) + \G(N_i\rightarrow \bar{l}\bar{\f})} \; .
\eeq 
To leading order in $\l^2$, the asymmetry is a sum of two terms, a mixing
contribution ($K=\l^{\dagger}\l$),
\beq\label{cpwave}
\e_i^M = - {1\over 8\p}\sum_{j\neq i} {M_i M_j \over M_i^2 - M_j^2}
        {\mbox{Im}\{K_{ij}^2\}\over K_{ii}}\; ,
\eeq
which is directly related to the one-loop self energy \cite{crv96,fps95},
and the familiar vertex contribution 
\beq\label{cpvertex}
\e_i^V =-{1\over 8\p}\sum_j {\mbox{Im}\{K_{ij}^2\}\over K_{ii}} 
         f\left({M_j^2\over M_i^2}\right) \; ,
\eeq
where
\beq
f(x) = \sqrt{x}\left(1-(1+x)\ln\left({1+x\over x}\right)\right)\;.
\eeq 
These results hold for sufficiently large mass splittings, i.e.
$|M_i-M_j|\gg |\G_i-\G_j|$. For small mass differences one expects an
enhancement of the mixing contribution \cite{fps96}. At present, however, the
influence of the thermal bath on this enhancement is unclear.

The use of classical Boltzmann equations with collision terms given by $S$-matrix
elements which crucially involve quantum interferences is unsatisfactory. Like
the collision terms also the time evolution of the system should be treated
quantum mechanically.

\section{Quantum mechanics of baryogenesis}

One would like to have a full quantum mechanical treatment
of baryogenesis. Starting with a density matrix with no initial 
asymmetry the emergence of an asymmetry should be seen from
the full time evolution. The Boltzmann equations should then follow
in some limit as a first-order approximation.

The Boltzmann equations are an on-shell approximation: 
between the interaction processes the particles propagate on-shell,
and for the scattering processes the on-shell $S$-matrix elements
are used. A full quantum treatment will also take off-shell effects into
account.

Of course, it is hard to obtain a full quantum description of the 
processes as the whole system is out of equilibrium. We shall therefore discuss
a toy model for a relaxation process, following Joichi, Matsumoto and 
Yoshimura \cite{jmy98}, and compare the exact description with the Boltzmann 
approach.

The model consists of a single quantum mechanical oscillator
coupled to a thermal bath of oscillators. The evolution of the system 
is governed by the Hamiltonian
\begin{eqnarray}
H & = & E c^{\dagger}c + \int^\infty_{\omega_c}\!\!\!\! d\omega\!\int\!\! d\Omega\:\,\omega 
b^{\dagger}(\Omega,\omega)b(\Omega,\omega) \\ && +\int^{\infty}_{\omega_c}\!\!\!\! d\omega \!\int\!\! d\Omega\: \left(
\zeta(\Omega,\omega) b^{\dagger}(\Omega,\omega)c + \mbox{ h.c.}
\right) \; .
\end{eqnarray}
$E$ is the frequency of the single oscillator, the frequencies of the
bath oscillators $\omega$ are bounded below by $\omega_c$ with 
$E>\omega_c$. $\Omega$ is an additional discrete or continuous label
for the bath oscillators, and $\zeta(\Omega,\omega)$ denotes the
complex coupling.

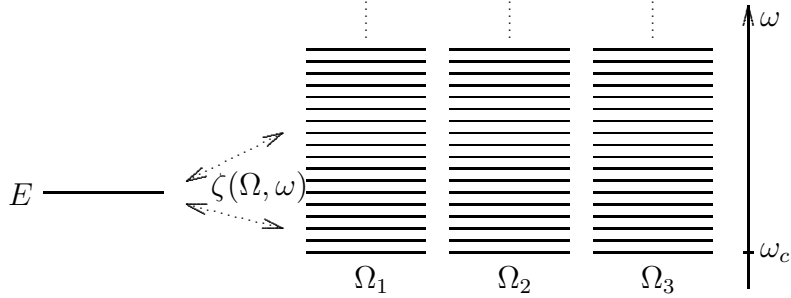
\begin{figure}
\begin{center}
\input{modfig}
\caption{\it One single oscillator with frequency $E$ is coupled
to a large thermal bath of oscillators.}
\end{center}
\end{figure}

The operators $c$ and $b$ fulfil canonical commutation relations,
\begin{eqnarray}
& [c,c^{\dagger}\,]=1 \quad , 
\quad  [c,b^{\dagger}(\Omega,\omega)]=0 \;,&\\
&[b(\Omega,\omega),b^{\dagger}(\Omega',\omega')]=\delta(\Omega-\Omega')
\delta(\omega-\omega') \;.&
\end{eqnarray}
The Hamiltonian is quadratic in $c$ and $b(\Omega,\omega)$ and 
hence solvable by a 
change of variables. It is possible to give an explicit formula for
new operators $B(\Omega,\omega)$ which diagonalize the Hamiltonian,
\begin{equation} 
H=\int_{\omega_c}^{\infty}\!\!\!\! d \omega \!\int\!\! d\Omega \:\, \omega 
B^{\dagger}(\Omega,\omega)B(\Omega,\omega) \; .
\end{equation}
These operators $B$ are obtained as a linear 
combination of $c$ and $b$. For $\omega$ away from $E$ it takes the
form
\begin{equation}
  B(\Omega,\omega)=b(\Omega,\omega)+\mathcal{O}(\zeta c,\zeta^* \zeta b)
\; .
\end{equation}
The time evolution in the new variables is just the free one,
\begin{equation}
  B^{\dagger}(\Omega,\omega,t)=e^{i\omega t}B^{\dagger}(\Omega,\omega)
\; .
\end{equation}
By inverting the change of variables one obtains explicit formulas
for the time evolution of
$c$ and $b$. Hence we are able to discuss exactly the properties
of the system. Let us start with a simple example.\\
\smallskip

\noindent\textbf{Decay process}\\

Assume that the system is prepared in the initial state $|\psi \rangle
= c^{\dagger}|0\rangle$ where only the single oscillator is excited.
We expect that this excited state decays. This should be seen from
the time evolution of the occupation number of the
single oscillator which is
\begin{equation}
 \langle \psi | c^{\dagger}(t)c(t)|\psi\rangle = |g(t)|^2\; , 
\end{equation}
where
\begin{equation}
g(t)=\int d\omega \sigma(\omega)\frac{1}{
(-\omega+E-\Pi(\omega))^2+(\pi\sigma(\omega))^2} e^{i\omega t} \;.
\end{equation}
Here $\sigma(\omega)$ is the absolute value squared
of the coupling summed over the internal label $\Omega$,
\begin{equation}
  \sigma(\omega)=\int\!\!d \Omega |\zeta(\Omega,\omega)|^2 \;.
\end{equation}
We see that $g$ is essentially the Fourier transform of the 
$c$-propagator
including the "self energy" $\Pi(\omega)+i\pi\sigma(\omega)$. 
For weak coupling, i.e. $\Pi(E)\ll E$ and $\sigma(E)\ll E-\omega_c$,
the integrand in the expression for $g$ has a sharp Breit-Wigner 
resonance at $\omega=E$. The contribution from this resonance is
\begin{equation}
  g_0(t)\propto e^{-\Gamma t/2+iEt} \;,
\end{equation}
where $\Gamma=2\pi\sigma(E)$. This is the result we would expect from the
Boltzmann equations: exponential decay with decay rate $\Gamma$. But 
this is not the only contribution. There is a second contribution
from the threshold which at large times only decreases
with a power law,
\begin{equation}
  g_1(t)\propto \kappa\frac{\Gamma(\alpha+1)}{(E-\omega_c)^2}
\frac{1}{t^{\alpha+1}} e^{i(\omega_c t+\alpha \pi/2)}\;.
\end{equation}
The constants $\kappa, \alpha$ parameterize the threshold behaviour
for $\omega$ close to $\omega_c$,
\begin{equation}
  \sigma(\omega)=\kappa (\omega-\omega_c)^{\alpha} \;.  
\end{equation}

At large times this contribution dominates the resonance contribution.
So we are led to the interesting result that the exponential 
Boltzmann decay is only relevant at intermediate times,
whereas the asymptotic behaviour is described by a power law.
This behaviour holds generally in field theory \cite{abv98}.\\
\smallskip

\noindent\textbf{Thermal equilibrium}\\

We will now study another interesting case, the situation of thermal
equilibrium. The density matrix is then given by
\begin{equation}
  \rho=e^{-\beta H} \quad,\quad \beta=\frac{1}{T} \;,
\end{equation}
which leads to the usual Bose-Einstein distribution in the variables
$B$,
\begin{equation}
  \langle B^{\dagger}(\Omega,\omega)B(\Omega',\omega')\rangle=
\delta(\Omega-\Omega')\delta(\omega-\omega')\frac{1}{e^{\beta\omega}-1}
\;.
\end{equation}
By expressing $c$ in terms of the operators $B$ we obtain the expectation
value of the occupation number of the single oscillator,
\begin{equation}
  \langle c^{\dagger}c\rangle=\int_{\omega_c}\!\!\!\!d\omega\sigma(\omega)
\frac{1}{
(-\omega+E-\Pi(\omega))^2+(\pi\sigma(\omega))^2}
\frac{1}{e^{\beta\omega}-1}\;.
\end{equation}
For weak coupling we have again a sharp resonance near $\omega=E$. 
The contribution from this resonance to the occupation number is
\begin{equation}
  \langle c^{\dagger}c\rangle \approx \frac{1}{e^{\beta E}-1}\;.
\end{equation}
This is the contribution we would expect for a free oscillator. However,
at small temperatures the resonance contribution is exponentially
suppressed and again the threshold behaviour becomes more important, yielding
\[
\langle c^{\dagger}c\rangle \approx \left\{ 
\begin{array}{r@{\quad,\quad\mbox{for}\quad}l}
\displaystyle
\kappa\frac{\zeta(\alpha+1)\Gamma(\alpha+1)}{(E-\omega_c)^2} 
T^{\alpha+1} & \omega_c\ll T\ll E \;,\\
\displaystyle
\kappa\frac{\Gamma(\alpha+1)}{(E-\omega_c)^2} 
e^{-\beta\omega_c}T^{\alpha+1} & T\ll \omega_c <E \;.
\end{array}
\right.
\]
Again we obtain a surprizing result: at small temperatures
the suppression of the occupation number is much weaker than what
is expected from the Bose-Einstein distribution.

What are the implications of this result? In \cite{my99} it has been argued 
that the power behaviour significantly affects the WIMP abundance.
However, the quantitative importance of this effect requires further
investigations \cite{ss99,yo99}.

What are the implications for baryogenesis and leptogenesis?
How large are the errors for the results 
obtained with standard Boltzmann
equations? To answer this question we try to apply our simple model
to the case of heavy particle decay.
For simplicity we consider only scalar particles. Let $X$ be a
heavy particle that may decay into light scalar particles $a$ and $b$.
The idea is to identify the heavy particle $X$ with the 
single oscillator and the decay products $a$ and $b$ with the
thermal oscillator bath.
The interaction is given by a Yukawa like coupling between the three 
scalar fields,
$H_{\rm int}=\lambda \int\!\!d^3 x \phi_X\phi_a\phi_b$ .

This interaction part should have the same structure as
the interaction part of the oscillator model. To see how this 
identification works we expand the fields in Fourier modes. As in the 
oscillator model we enforce that total particle number is conserved, i.e. we 
only take into account those parts of the Hamiltonian which describe decay and 
inverse decay,
\begin{eqnarray}
  H_{\rm int}& =& \lambda \int\!\!d^3 x \phi_X\phi_a\phi_b \nonumber\\
& \rightarrow & \lambda\int\!\! 
\frac{d^3q}{(2\pi)^3(2\omega_q)^{1/2}}
\frac{d^3k_a}{(2\pi)^3(2\omega_a)^{1/2}}
\frac{d^3k_b}{(2\pi)^3(2\omega_b)^{1/2}}(2\pi)^3
\delta^{(3)}(\vec{q}-\vec{k_a}-\vec{k_b})\nonumber\\
&&\quad\left(c^{\dagger}
(\vec{q}\,)b_a(\vec{k_a})b_b(\vec{k_b}) 
+\mbox{h.c.} \right) \nonumber \\
\label{identifizierung}
& \stackrel{!}{=} & \int\!\! 
\frac{d^3q}{(2\pi)^3}\int\!\! d\omega d\Omega 
\zeta(\Omega,\omega,\vec{q}\,)
\left(c^{\dagger}(\vec{q})b(\Omega,\omega,\vec{q}\,) 
+\mbox{h.c.}\right) \; .
\end{eqnarray}
One then reads off the definition of composite operators,
\begin{equation}
  b^{\dagger}(\Omega,\omega,\vec{q}\,)=\bigg(
\frac{1}{(2\pi)^3}\bigg| 
\frac{\partial(\vec{k_a},\vec{k_b})}{\partial(\Omega,\omega,\vec{q}\,)} 
\bigg|\bigg)^{1/2}b_a^{\dagger}
(\vec{k_a})b_b^{\dagger}(\vec{k_b}) 
\; ,
\end{equation}
which are needed for the identification. Here $\Omega$ 
is a label
that together with the energy $\omega$ and the total momentum $\vec{q}$
determines the momenta $\vec{k_a},\vec{k_b}$, e.\,g. it can be chosen
as two angles describing the direction of $\vec{k_a}$. 

These composite operators fulfil commutation
relations that differ from the canonical ones. To apply the oscillator
model we try to approximate the commutation relation and the
kinetic term of the thermal bath as
\begin{equation}
  [b(\Omega,\omega,\vec{q}\,),
        b^{\dagger}(\Omega',\omega',\vec{q'}\,)]
        = (2\pi)^3\delta(\vec{q}-\vec{q'})
         \delta(\omega-\omega')\delta(\Omega-\Omega') \;,
\end{equation}
\begin{equation}
  H_{\rm bath}=\int\!\!\frac{d^3q}{(2\pi)^3}
d\omega d\Omega \, \omega 
b^{\dagger}(\Omega,\omega,\vec{q}\,)b(\Omega,\omega,\vec{q}\,)\;.
\end{equation}
With these adjustments one can compute the decay of the heavy particle
$X$. The approximations made above do not affect the behaviour of
$X$ substantially and correspond to a low density approximation.
Again we obtain the result of an exponential decay followed by a 
power law behaviour at large times.

When describing properties of the thermal bath like the time
evolution of an asymmetry things become more involved and it is
not possible to describe the long-term behaviour in the model.
But nevertheless it appears possible to compute the time derivative
of the asymmetry correctly. 

An important question is how to implement $C\!P$ violation
in the oscillator model without breaking $C\!PT$ invariance.
This can only be done in an extended version of the oscillator model which 
involves an extra interaction term. In connection with leptogenesis this 
corresponds to effects of the heavy neutrinos $N_2$ and $N_3$.

\begin{figure}[h]
\begin{center}
\epsfig{file=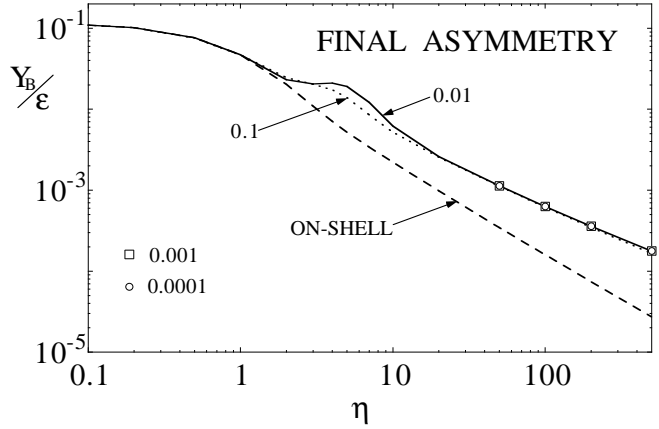,height=6cm,clip=,bbllx=97,bblly=285,bburx=472,bbury=535}
\caption{\label{fig:asymmetry}\it The final asymmetry as function of
$\eta=\frac{\Gamma}{H}\Big|_{T=M}$ for two values of
$\frac{\Gamma}{M}$: $0.0001$ (sharp resonance) and $0.1$ 
(broad resonance).}
\end{center}
\end{figure}

The production rate of the asymmetry can now be calculated. The 
resonance contribution coincides exactly with the rate given in 
section 2 including both contributions (\ref{cpwave}) and (\ref{cpvertex}) 
to the $C\!P$ asymmetry. 
A detailed analysis of these results is given in \cite{fr99}.

Despite the various remaining problems it is worthwhile to 
estimate the size of the corrections for the baryogenesis scenario. 
In \cite{jmy98} such an estimate is given under certain assumptions and 
approximations. The result is shown in fig.~\ref{fig:asymmetry}. If the 
decay rate $\Gamma$ is of order  the
Hubble rate $H$ or smaller there is no significant deviation from the 
result obtained with the Boltzmann equations (on-shell result). For
$\Gamma/H>1$, where the asymmetry is smaller because the system is
closer to equilibrium, the asymmetry production is enhanced by
off-shell effects.

\section{Conclusions}

At present there still exists a variety of mechanisms which, at least
in principle, can account for the cosmological baryon asymmetry. 
Particularly successful is the leptogenesis scenario. Given the experimental
indications for neutrino masses it naturally explains the observed order
of magnitude of the cosmological baryon asymmetry without any fine tuning 
of parameters.

One is therefore led to examine the theoretical basis for current estimates
of the baryon asymmetry. Although our understanding of the high-temperature
symmetric phase of the standard model has significantly improved in recent
years the quantitative description of an out-of-equilibrium process
remains a difficult problem.

Particularly subtle is the use of the classical Boltzmann equations together
with collision terms derived from $S$-matrix elements which involve
quantum interferences in a crucial manner. A full quantum mechanical
treatment which includes the time evolution of the system is highly
desirable. At present several interesting ideas are pursued by different
groups but a fully satisfactory solution of the problem still remains to
be found.

\end{document}

%% file: Fig01.tex
 \begin{center}
     \pspicture*(-0.50,-2.5)(8.5,6.5)
     \psset{linecolor=lightgray}
     \qdisk(4,2){1.5cm}
     \psset{linecolor=black}
     \pscircle[linewidth=1pt,linestyle=dashed](4,2){1.5cm}
     \rput[cc]{0}(4,2){\scalebox{1.5}{Sphaleron}}
     \psline[linewidth=1pt](5.50,2.00)(7.50,2.00)
     \psline[linewidth=1pt](5.30,2.75)(7.03,3.75)
     \psline[linewidth=1pt](4.75,3.30)(5.75,5.03)
     \psline[linewidth=1pt](4.00,3.50)(4.00,5.50)
     \psline[linewidth=1pt](3.25,3.30)(2.25,5.03)
     \psline[linewidth=1pt](2.70,2.75)(0.97,3.75)
     \psline[linewidth=1pt](2.50,2.00)(0.50,2.00)
     \psline[linewidth=1pt](2.70,1.25)(0.97,0.25)
     \psline[linewidth=1pt](3.25,0.70)(2.25,-1.03)
     \psline[linewidth=1pt](4.00,0.50)(4.00,-1.50)
     \psline[linewidth=1pt](4.75,0.70)(5.75,-1.03)
     \psline[linewidth=1pt](5.30,1.25)(7.03,0.25)
     \psline[linewidth=1pt]{<-}(6.50,2.00)(6.60,2.00)
     \psline[linewidth=1pt]{<-}(6.17,3.25)(6.25,3.30)
     \psline[linewidth=1pt]{<-}(5.25,4.17)(5.30,4.25)
     \psline[linewidth=1pt]{<-}(4.00,4.50)(4.00,4.60)
     \psline[linewidth=1pt]{<-}(2.75,4.17)(2.70,4.25)
     \psline[linewidth=1pt]{<-}(1.83,3.25)(1.75,3.30)
     \psline[linewidth=1pt]{<-}(1.50,2.00)(1.40,2.00)
     \psline[linewidth=1pt]{<-}(1.83,0.75)(1.75,0.70)
     \psline[linewidth=1pt]{<-}(2.75,-0.17)(2.70,-0.25)
     \psline[linewidth=1pt]{<-}(4.00,-0.50)(4.00,-0.60)
     \psline[linewidth=1pt]{<-}(5.25,-0.17)(5.30,-0.25)
     \psline[linewidth=1pt]{<-}(6.17,0.75)(6.25,0.70)
     \rput[cc]{0}(8.00,2.00){\scalebox{1.3}{$b_L$}}
     \rput[cc]{0}(7.46,4.00){\scalebox{1.3}{$b_L$}}
     \rput[cc]{0}(6.00,5.46){\scalebox{1.3}{$t_L$}}
     \rput[cc]{0}(4.00,6.00){\scalebox{1.3}{$s_L$}}
     \rput[cc]{0}(2.00,5.46){\scalebox{1.3}{$s_L$}}
     \rput[cc]{0}(0.54,4.00){\scalebox{1.3}{$c_L$}}
     \rput[cc]{0}(0.00,2.00){\scalebox{1.3}{$d_L$}}
     \rput[cc]{0}(0.54,0.00){\scalebox{1.3}{$d_L$}}
     \rput[cc]{0}(2.00,-1.46){\scalebox{1.3}{$u_L$}}
     \rput[cc]{0}(4.00,-2.00){\scalebox{1.3}{$\nu_e$}}
     \rput[cc]{0}(6.00,-1.46){\scalebox{1.3}{$\nu_{\mu}$}}
     \rput[cc]{0}(7.46,0.00){\scalebox{1.3}{$\nu_{\tau}$}}
     \endpspicture
\end{center}

%% file: Eff_Interaction.tex
 \begin{center}
     \pspicture*(-0.50,-0.25)(8.5,4.5)
     \psline[linewidth=1pt](4,2)(7.03,3.75)
     \psline[linewidth=1pt](4,2)(0.97,3.75)
     \psline[linewidth=1pt,linestyle=dashed](4,2)(0.97,0.25)
     \psline[linewidth=1pt,linestyle=dashed](4,2)(7.03,0.25)
     \psset{linecolor=lightgray}
     \qdisk(4,2){1.2cm}
     \psset{linecolor=black}
     \pscircle[linewidth=1pt](4,2){1.2cm}
     \psline[linewidth=2pt]{<-}(6.17,3.25)(6.25,3.30)
     \psline[linewidth=2pt]{<-}(1.83,3.25)(1.75,3.30)
     \psline[linewidth=2pt]{<-}(1.83,0.75)(1.75,0.70)
     \psline[linewidth=2pt]{<-}(6.17,0.75)(6.25,0.70)
     \rput[cc]{0}(7.46,4.00){\scalebox{1.3}{$\bar{l}_L$}}
     \rput[cc]{0}(0.54,4.00){\scalebox{1.3}{$l_L$}}
     \rput[cc]{0}(0.54,0.00){\scalebox{1.3}{$\phi$}}
     \rput[cc]{0}(7.46,0.00){\scalebox{1.3}{$\bar{\phi}$}}
     \endpspicture
\end{center}

%% file: decay1.tex
 \begin{center}
     \pspicture*(-0.5,-0.5)(8.5,4.5)
     \psset{linecolor=lightgray}
     \qdisk(4,2){1cm}
     \psset{linecolor=black}
     \pscircle[linewidth=1pt](4,2){1cm}
     \psline[linewidth=1pt](4.9,2.5)(7.03,3.75)
     \psline[linewidth=1pt](3.00,2.00)(0.90,2.00)
     \psline[linewidth=1pt,linestyle=dashed](4.9,1.5)(7.03,0.25)
     \psline[linewidth=2pt]{->}(6.17,3.25)(6.25,3.30)
     \psline[linewidth=2pt]{->}(6.17,0.75)(6.25,0.70)
     \rput[cc]{0}(7.46,4.00){\scalebox{1.6}{$l$}}
     \rput[cc]{0}(0.40,2.00){\scalebox{1.6}{$N$}}
     \rput[cc]{0}(7.46,0.00){\scalebox{1.6}{$\phi$}}
     \endpspicture
\end{center}

%% file: decay2.tex
 \begin{center}
     \pspicture*(-0.50,-0.5)(8.5,4.5)
     \psset{linecolor=lightgray}
     \qdisk(4,2){1cm}
     \psset{linecolor=black}
     \pscircle[linewidth=1pt](4,2){1cm}
     \psline[linewidth=1pt](4.9,2.5)(7.03,3.75)
     \psline[linewidth=1pt](3.00,2.00)(0.90,2.00)
     \psline[linewidth=1pt,linestyle=dashed](4.9,1.5)(7.03,0.25)
     \psline[linewidth=2pt]{<-}(6.17,3.25)(6.25,3.30)
     \psline[linewidth=2pt]{<-}(6.17,0.75)(6.25,0.70)
     \rput[cc]{0}(7.46,4.00){\scalebox{1.6}{$\bar{l}$}}
     \rput[cc]{0}(0.40,2.00){\scalebox{1.6}{$N$}}
     \rput[cc]{0}(7.46,0.00){\scalebox{1.6}{$\bar{\phi}$}}
     \endpspicture
\end{center}

%% file: decay3.tex
 \begin{center}
     \pspicture*(-0.50,-0.5)(8.5,4.5)
     \psset{linecolor=lightgray}
     \qdisk(4,2){1cm}
     \psset{linecolor=black}
     \pscircle[linewidth=1pt](4,2){1cm}
     \psline[linewidth=1pt](5.0,2.00)(7.10,2.00)
     \psline[linewidth=1pt](3.13,2.50)(0.97,3.75)
     \psline[linewidth=1pt,linestyle=dashed](3.13,1.50)(0.97,0.25)
     \psline[linewidth=2pt]{<-}(1.83,3.25)(1.75,3.30)
     \psline[linewidth=2pt,linestyle=dashed]{<-}(1.83,0.75)(1.75,0.70)
     \rput[cc]{0}(7.60,2.00){\scalebox{1.6}{$N$}}
     \rput[cc]{0}(0.54,4.00){\scalebox{1.6}{$l$}}
     \rput[cc]{0}(0.54,0.00){\scalebox{1.6}{$\phi$}}
     \endpspicture
\end{center}

%% file: decay4.tex
 \begin{center}
     \pspicture*(-0.50,-0.5)(8.5,4.5)
     \psset{linecolor=lightgray}
     \qdisk(4,2){1cm}
     \psset{linecolor=black}
     \pscircle[linewidth=1pt](4,2){1cm}
     \psline[linewidth=1pt](5.0,2.00)(7.10,2.00)
     \psline[linewidth=1pt](3.13,2.50)(0.97,3.75)
     \psline[linewidth=1pt,linestyle=dashed](3.13,1.50)(0.97,0.25)
     \psline[linewidth=2pt]{->}(1.83,3.25)(1.75,3.30)
     \psline[linewidth=2pt,linestyle=dashed]{->}(1.83,0.75)(1.75,0.70)
     \rput[cc]{0}(7.60,2.00){\scalebox{1.6}{$N$}}
     \rput[cc]{0}(0.54,4.00){\scalebox{1.6}{$\bar{l}$}}
     \rput[cc]{0}(0.54,0.00){\scalebox{1.6}{$\bar{\phi}$}}
     \endpspicture
\end{center}

%% file: twotwo.tex
 \begin{center}
     \pspicture*(-0.50,-0.5)(8.5,4.5)
     \psline[linewidth=1pt](4,2)(7.03,3.75)
     \psline[linewidth=1pt](4,2)(0.97,3.75)
     \psline[linewidth=1pt,linestyle=dashed](4,2)(0.97,0.25)
     \psline[linewidth=1pt,linestyle=dashed](4,2)(7.03,0.25)
     \psset{linecolor=lightgray}
     \qdisk(4,2){1cm}
     \psset{linecolor=black}
     \pscircle[linewidth=1pt](4,2){1cm}
     \psline[linewidth=2pt]{->}(6.17,3.25)(6.25,3.30)
     \psline[linewidth=2pt]{<-}(1.83,3.25)(1.75,3.30)
     \psline[linewidth=2pt]{<-}(1.83,0.75)(1.75,0.70)
     \psline[linewidth=2pt]{->}(6.17,0.75)(6.25,0.70)
     \rput[cc]{0}(7.46,4.00){\scalebox{1.6}{$l$}}
     \rput[cc]{0}(0.54,4.00){\scalebox{1.6}{$l$}}
     \rput[cc]{0}(0.54,0.00){\scalebox{1.6}{$\phi$}}
     \rput[cc]{0}(7.46,0.00){\scalebox{1.6}{$\phi$}}
     \endpspicture
\end{center}

%% file: twoanti.tex
 \begin{center}
     \pspicture*(-0.50,-0.5)(8.5,4.5)
     \psline[linewidth=1pt](4,2)(7.03,3.75)
     \psline[linewidth=1pt](4,2)(0.97,3.75)
     \psline[linewidth=1pt,linestyle=dashed](4,2)(0.97,0.25)
     \psline[linewidth=1pt,linestyle=dashed](4,2)(7.03,0.25)
     \psset{linecolor=lightgray}
     \qdisk(4,2){1cm}
     \psset{linecolor=black}
     \pscircle[linewidth=1pt](4,2){1cm}
     \psline[linewidth=2pt]{<-}(6.17,3.25)(6.25,3.30)
     \psline[linewidth=2pt]{<-}(1.83,3.25)(1.75,3.30)
     \psline[linewidth=2pt]{<-}(1.83,0.75)(1.75,0.70)
     \psline[linewidth=2pt]{<-}(6.17,0.75)(6.25,0.70)
     \rput[cc]{0}(7.46,4.00){\scalebox{1.6}{$\bar{l}$}}
     \rput[cc]{0}(0.54,4.00){\scalebox{1.6}{$l$}}
     \rput[cc]{0}(0.54,0.00){\scalebox{1.6}{$\phi$}}
     \rput[cc]{0}(7.46,0.00){\scalebox{1.6}{$\bar{\phi}$}}
     \endpspicture
\end{center}

%% file: outofequ.tex
\font\thinlinefont=cmr5
\begingroup\makeatletter\ifx\SetFigFont\undefined
\def\x#1#2#3#4#5#6#7\relax{\def\x{#1#2#3#4#5#6}}%
\expandafter\x\fmtname xxxxxx\relax \def\y{splain}%
\ifx\x\y   
\gdef\SetFigFont#1#2#3{%
  \ifnum #1<17\tiny\else \ifnum #1<20\small\else
  \ifnum #1<24\normalsize\else \ifnum #1<29\large\else
  \ifnum #1<34\Large\else \ifnum #1<41\LARGE\else
     \huge\fi\fi\fi\fi\fi\fi
  \csname #3\endcsname}%
\else
\gdef\SetFigFont#1#2#3{\begingroup
  \count@#1\relax \ifnum 25<\count@\count@25\fi
  \def\x{\endgroup\@setsize\SetFigFont{#2pt}}%
  \expandafter\x
    \csname \romannumeral\the\count@ pt\expandafter\endcsname
    \csname @\romannumeral\the\count@ pt\endcsname
  \csname #3\endcsname}%
\fi
\fi\endgroup
\mbox{\beginpicture
\setcoordinatesystem units <0.020000cm,0.01700cm>
\unitlength=0.050000cm
\linethickness=1pt
\setplotsymbol ({\makebox(0,0)[l]{\tencirc\symbol{'160}}})
\setshadesymbol ({\thinlinefont .})
\setlinear
%
%
\linethickness= 0.500pt
\setplotsymbol ({\thinlinefont .})
\setdashes < 0.0952cm>
\plot 88.375 252.5 
  92.0625 252.5  
  96.1875 252.5
    100.438  252.5
    104.375  252.5
    108.812  252.5  
  112.812  252.5
    117.5  252.5
    121.875  252.5 
   125.938  252.5
    130.188  252.5
    133.812  252.5
    137.812  252.5
    142  252.5
    145.875  252.5
    150.312  252.5
    154.562  252.5
    158.5  252.5
    162.75  252.5 
   170.375  252.5
    174.688  252.5
    179.25  252.5
    183.625  252.5
    187.625  252.438 
   195.438  252.438
    199.875  252.438 
   204  252.375 
   208.375  252.375 
   212.5  252.312 
   220.188  252.25
    224.188  252.188 
   228.438  252.125  
  236.062  251.875
    240  251.75
    244.188  251.562
    251.75  251.125 
   255.75  250.812
    260.062  250.375
    264.625  249.812
    268.812  249.25
    276.938  247.75
    285.625  245.5
    293.625  242.562 302.25  238.25
    310.5  232.688 
   319.25  224.75 
   327.375  215 
   334.938  203.062 
   343.375  186.062 
   351.062  166  
   358.938  139.75
   367.562  102.875 
   375.75  57.625
   382.75  7.75 /

%
%
\linethickness= 0.500pt
\setplotsymbol ({\thinlinefont .})
\setsolid
\plot 88.375  252.5
    92.0625  252.5 
   96.1875  252.5
    100.438  252.5
    104.375  252.5 
   108.812  252.5 
   112.812  252.5 
   117.5  252.5  
  121.875  252.5 
   125.938  252.5 
   130.188  252.5  
  133.812  252.5 
   137.812  252.5 
   142  252.5  
  145.875  252.5  
  150.312  252.5 
   154.562  252.5   
 158.5  252.5  
  162.75  252.5
    166.75  252.5 
   170.375  252.5
    174.25  252.5 
   178.375  252.5  
  182.875  252.5  
  187  252.5   
 190.938  252.5
    195.188  252.5 
   199.125  252.5  
  202.75  252.5  
  206.625  252.5 
   210.688  252.438 
   215.188  252.438 
   219.312  252.438  
  223.188  252.438  
  227.375  252.375  
  235  252.312 
   238.938  252.312 
   243.312  252.25  
  247.812  252.125  
  252.062  252.062 
   256.438  251.938 
   260.562  251.75 
   268.25  251.312  
  272.25  251
    276.562  250.625   
 284.188  249.625  
  288.125  248.938 
   292.312  248 
   299.938  245.75
    304.125  244.125
 308.75  241.875 
313.125  239.25
 317.062  236.375 
325.25  228.625 
333.938  217.062 
341.938  202.5 
350.625  181.438
 358.875  154.688
367.625  117.625 
375.812  72.6875 
383.438  19.4375 
384.75  7.75 /

%
%
\linethickness= 0.500pt
\setplotsymbol ({\thinlinefont .})
\setsolid
\plot 206.875  7.75
    214.25  22.0625 
   222.625  34.625
    230.812  44.3125 
   238.438  51.375 
   246.25  57.1875
    254.688  62.0625  
  263.562  66  
  271.938  68.875 
   280.25  71.0625 
   289  72.875 
   297.125  74.125 
   304.75  75.0625 
   313.188  75.8125  
  320.875  76.375 
   328.812  76.75  
  337.438  77.125 
   345.625  77.3125  
  349.812  77.4375   
 354.312  77.5
    362.438  77.625 
   370  77.6875 
   374  77.75 
   378.312  77.75 
   382.875  77.75 
   387.062  77.8125 
   395.188  77.8125 
   399.688  77.8125  
  403.875  77.8125 
   407.938  77.8125 
   412.188  77.8125 
   419.875  77.875 
   423.812  77.875 
   428  77.875 
   435.562  77.875  
  439.812  77.875  
  444.375  77.875 
   448.688  77.875 
   452.688  77.875  
  457.125  77.875 
   461.25  77.875 
   465  77.875  
  469  77.875 
   472.625  77.875  
  476.562  77.875   
 483.562  77.875 /
%
%
\linethickness= 0.500pt
\setplotsymbol ({\thinlinefont .})
\setsolid
\plot 75 0 490 0 /
%
%
\linethickness= 0.500pt
\setplotsymbol ({\thinlinefont .})
\setsolid
\plot 75 270 75 0 /
%
%
\linethickness= 0.500pt
\setplotsymbol ({\thinlinefont .})
\setsolid
\plot 75 270 80 265 /
%
%
\linethickness= 0.500pt
\setplotsymbol ({\thinlinefont .})
\setsolid
\plot 75 270 70 265 /
%
%
\linethickness= 0.500pt
\setplotsymbol ({\thinlinefont .})
\setsolid
\plot 490 0 485 5 /
%
%
\linethickness= 0.500pt
\setplotsymbol ({\thinlinefont .})
\setsolid
\plot 490 0 485 -5 /
%
%
\linethickness= 0.500pt
\setplotsymbol ({\thinlinefont .})
\setsolid
\plot 246.438 0 246.438 5 /

%
%
\put{$T^{-1}$} [lB] at 500 -5
%
%
\put{$M^{-1}$} [lB] at 240 -25
%
%
\put{$Y_N$} [lB] at 350 220
%
%
\put{$Y_N^{eq}$} [lB] at 290 170
%
%
\put{$Y_l-Y_{\overline{l}}$} [lB] at 460 90
%
%
\put{$Y$} [lB] at 40 240
\linethickness=0pt
\putrectangle corners at 10 280 and 550 -25
\endpicture}

%% file: vertex.tex
\begin{center}
  \psset{unit=1.2cm}
  \pspicture[0.5](-2.2,0.5)(3.5,3.5)
      \psline[linewidth=1pt](-1.65,3.22)(-0.65,2)
      \psline[linewidth=1pt,linestyle=dashed](-1.65,0.78)(-0.65,2)
      \psline[linewidth=1pt](-0.65,2)(1.35,2)
      \psline[linewidth=2pt]{->}(-1.15,1.39)(-1.05,1.49)
      \psline[linewidth=2pt]{->}(-1.15,2.61)(-1.05,2.51)
      \rput[cc]{0}(0.35,2.3){$\displaystyle N$}
      \rput[cc]{0}(-1.95,3.22){$\displaystyle l$}
      \rput[cc]{0}(-1.95,0.78){$\displaystyle \phi$}
      \psarc[linewidth=1pt](1.95,2){0.6}{-180}{60}
      \psarc[linewidth=1pt,linestyle=dashed](1.95,2){0.6}{60}{180}
      \psline[linewidth=2pt]{<-}(1.69,1.46)(1.57,1.54)
      \psline[linewidth=2pt]{<-}(1.63,2.51)(1.54,2.43)
      \rput[cc]{0}(2.75,2){$\displaystyle N$}
      \rput[cc]{0}(1.5,2.78){$\displaystyle \phi$}
      \rput[cc]{0}(1.5,1.22){$\displaystyle l$}
      \psline[linewidth=1pt,linestyle=dashed](2.25,1.48)(2.95,0.78)
      \psline[linewidth=1pt](2.25,2.52)(2.95,3.22)
      \psline[linewidth=2pt]{<-}(2.6,2.87)(2.7,2.97)
      \psline[linewidth=2pt]{<-}(2.53,1.2)(2.63,1.1)
      \rput[cc]{0}(3.3,3.22){$\displaystyle \bar{l}$}
      \rput[cc]{0}(3.3,0.78){$\displaystyle \bar{\phi}$}
  \endpspicture
\end{center}

%% file: selfen.tex
\begin{center}
    \pspicture(-1,0)(8,3.5)
      \psline[linewidth=1pt](-0.4,2.7)(0.6,1.5)
      \psline[linewidth=1pt,linestyle=dashed](-0.4,0.3)(0.6,1.5)
      \psline[linewidth=2pt]{->}(0.1,2.1)(0.2,2.0)
      \psline[linewidth=2pt]{->}(0.1,0.9)(0.2,1.0)
      \rput[cc]{0}(-0.7,2.7){$\displaystyle l$}
      \rput[cc]{0}(-0.7,0.6){$\displaystyle \phi$}
      \psline[linewidth=1pt](0.6,1.5)(1.8,1.5)
      \psline[linewidth=1pt](3.4,1.5)(4.6,1.5)
      \rput[cc]{0}(1.1,1.9){$N_i$}
      \rput[cc]{0}(3.9,1.9){$N_j$}
      \psarc[linewidth=1pt](2.6,1.5){0.8}{-180}{0}
      \psarc[linewidth=1pt,linestyle=dashed](2.6,1.5){0.8}{0}{180}
      \psline[linewidth=2pt]{->}(2.62,2.29)(2.72,2.29)
      \psline[linewidth=2pt]{->}(2.62,0.71)(2.72,0.71)
      \rput[cc]{0}(2.6,0.3){$l$}
      \rput[cc]{0}(2.6,2.7){$\phi$}
      \psline[linewidth=1pt](4.6,1.5)(5.6,2.7)
      \psline[linewidth=1pt,linestyle=dashed](4.6,1.5)(5.6,0.3)
      \psline[linewidth=2pt]{<-}(5.1,2.1)(5.2,2.2)
      \psline[linewidth=2pt]{<-}(5.1,0.9)(5.2,0.8)
      \rput[cc]{0}(5.9,2.7){$\displaystyle \bar{l}$}
      \rput[cc]{0}(5.9,0.3){$\displaystyle \bar{\phi}$}
    \endpspicture
\end{center}

%% file: prop.tex
\begin{center}
\scaleboxto(6.3,0){\parbox{7cm}{
  \pspicture[0.5](0.5,0.65)(7.5,3.35)
      \psline[linewidth=1pt](1.0,3.05)(1.0,0.95)
      \psline[linewidth=1pt,linestyle=dashed](1.6,0.95)(2.65,2)
      \psline[linewidth=1pt](1.6,3.05)(2.65,2)
      \psline[linewidth=1pt]{->}(1.82,1.18)(1.92,1.28)
      \psline[linewidth=1pt]{->}(1.82,2.83)(1.92,2.73)
      \rput[cc]{0}(1.3,0.95){\scalebox{1.2}{$\displaystyle \phi$}}
      \rput[cc]{0}(1.3,3.05){\scalebox{1.2}{$\displaystyle l$}}
      \psline[linewidth=1pt](2.65,2)(3.5,2)
      \pscircle*[linewidth=1pt,fillstyle=solid,
                 linecolor=lightgray](4,2){0.5}
      \pscircle[linewidth=1pt,fillstyle=none](4,2){0.5}
      \psline[linewidth=1pt](4.5,2)(5.35,2)
      \rput[cc]{0}(4,2.8){\scalebox{1.2}{$\displaystyle N$}}
      \psline[linewidth=1pt](5.35,2)(6.4,3.05)
      \psline[linewidth=1pt,linestyle=dashed](5.35,2)(6.4,0.95)
      \psline[linewidth=1pt]{->}(6.18,2.83)(6.08,2.73)
      \psline[linewidth=1pt]{->}(6.18,1.18)(6.08,1.28)
      \rput[cc]{0}(6.7,3.05){\scalebox{1.2}{$\displaystyle \bar{l}$}}
      \rput[cc]{0}(6.7,0.95){\scalebox{1.2}{$\displaystyle \bar{\phi}$}}
      \psline[linewidth=1pt](7.0,3.05)(7.0,0.95)
      \rput[cc]{0}(7.3,3.05){$\displaystyle 2$}
  \endpspicture}}
  $\quad - \quad$
\scaleboxto(6.3,0){\parbox{7cm}{
  \pspicture[0.5](0.5,0.65)(7.5,3.35)
      \psline[linewidth=1pt](1.0,3.05)(1.0,0.95)
      \psline[linewidth=1pt,linestyle=dashed](1.6,0.95)(2.65,2)
      \psline[linewidth=1pt](1.6,3.05)(2.65,2)
      \psline[linewidth=1pt]{<-}(1.82,1.18)(1.92,1.28)
      \psline[linewidth=1pt]{<-}(1.82,2.83)(1.92,2.73)
      \rput[cc]{0}(1.3,0.95){\scalebox{1.2}{$\displaystyle \bar{\phi}$}}
      \rput[cc]{0}(1.3,3.05){\scalebox{1.2}{$\displaystyle \bar{l}$}}
      \psline[linewidth=1pt](2.65,2)(3.5,2)
      \pscircle*[linewidth=1pt,fillstyle=solid,
                 linecolor=lightgray](4,2){0.5}
      \pscircle[linewidth=1pt,fillstyle=none](4,2){0.5}
      \psline[linewidth=1pt](4.5,2)(5.35,2)
      \rput[cc]{0}(4,2.8){\scalebox{1.2}{$\displaystyle N$}}
      \psline[linewidth=1pt](5.35,2)(6.4,3.05)
      \psline[linewidth=1pt,linestyle=dashed](5.35,2)(6.4,0.95)
      \psline[linewidth=1pt]{<-}(6.24,2.89)(6.14,2.79)
      \psline[linewidth=1pt]{<-}(6.24,1.11)(6.14,1.21)
      \rput[cc]{0}(6.7,3.05){\scalebox{1.2}{$\displaystyle l$}}
      \rput[cc]{0}(6.7,0.95){\scalebox{1.2}{$\displaystyle \phi$}}
      \psline[linewidth=1pt](7.0,3.05)(7.0,0.95)
      \rput[cc]{0}(7.3,3.05){$\displaystyle 2$} 
  \endpspicture}}
$=\quad 0$
\end{center}

%% file: full.tex
\begin{center}
  \pspicture[0.5](0.5,0.65)(5.0,3.35)
      \psline[linewidth=1pt](1.0,3.05)(1.0,0.95)
      \psline[linewidth=1pt,linestyle=dashed](1.6,0.95)(2.3,1.65)
      \psline[linewidth=1pt](1.6,3.05)(2.3,2.35)
      \psline[linewidth=1pt]{->}(1.82,1.18)(1.92,1.28)
      \psline[linewidth=1pt]{->}(1.82,2.83)(1.92,2.73)
      \rput[cc]{0}(1.3,0.95){$\displaystyle \phi$}
      \rput[cc]{0}(1.3,3.05){$\displaystyle l$}
      \pscircle*[linewidth=1pt,fillstyle=solid,
                 linecolor=lightgray](2.65,2){0.5}
      \pscircle[linewidth=1pt,fillstyle=none](2.65,2){0.5}
      \psline[linewidth=1pt](3.0,2.35)(3.7,3.05)
      \psline[linewidth=1pt,linestyle=dashed](3.0,1.65)(3.7,0.95)
      \psline[linewidth=1pt]{->}(3.48,2.83)(3.38,2.73)
      \psline[linewidth=1pt]{->}(3.48,1.18)(3.38,1.28)
      \rput[cc]{0}(4.0,3.05){$\displaystyle \bar{l}$}
      \rput[cc]{0}(4.0,0.95){$\displaystyle \bar{\phi}$}
      \psline[linewidth=1pt](4.3,3.05)(4.3,0.95)
      \rput[cc]{0}(4.6,3.05){$\displaystyle 2$}
  \endpspicture
  $\quad - \quad$
  \pspicture[0.5](0.5,0.65)(5.0,3.35)
      \psline[linewidth=1pt](1.0,3.05)(1.0,0.95)
      \psline[linewidth=1pt,linestyle=dashed](1.6,0.95)(2.3,1.65)
      \psline[linewidth=1pt](1.6,3.05)(2.3,2.35)
      \psline[linewidth=1pt]{<-}(1.82,1.18)(1.92,1.28)
      \psline[linewidth=1pt]{<-}(1.82,2.83)(1.92,2.73)
      \rput[cc]{0}(1.3,0.95){$\displaystyle \bar{\phi}$}
      \rput[cc]{0}(1.3,3.05){$\displaystyle \bar{l}$}
      \pscircle*[linewidth=1pt,fillstyle=solid,
                 linecolor=lightgray](2.65,2){0.5}
      \pscircle[linewidth=1pt,fillstyle=none](2.65,2){0.5}
      \psline[linewidth=1pt](3,2.35)(3.7,3.05)
      \psline[linewidth=1pt,linestyle=dashed](3,1.65)(3.7,0.95)
      \psline[linewidth=1pt]{<-}(3.54,2.89)(3.44,2.79)
      \psline[linewidth=1pt]{<-}(3.54,1.11)(3.44,1.21)
      \rput[cc]{0}(4,3.05){$\displaystyle l$}
      \rput[cc]{0}(4,0.95){$\displaystyle \phi$}
      \psline[linewidth=1pt](4.3,3.05)(4.3,0.95)
      \rput[cc]{0}(4.6,3.05){$\displaystyle 2$}
  \endpspicture
  $\quad = \quad 0\;$ .
\end{center}

%% file: fullv.tex
\begin{center}
  \pspicture[0.5](4,0.65)(8.0,3.35)
      \psline[linewidth=1pt](5.0,2)(5.85,2)
      \rput[cc]{0}(4.7,2){$\displaystyle N$}
      \pscircle*[linewidth=1pt,fillstyle=solid,
                 linecolor=lightgray](6.35,2){0.5}
      \pscircle[linewidth=1pt,fillstyle=none](6.35,2){0.5}
      \psline[linewidth=1pt](6.7,2.35)(7.4,3.05)
      \psline[linewidth=1pt,linestyle=dashed](6.7,1.65)(7.4,0.95)
      \psline[linewidth=2pt]{<-}(7.24,2.89)(7.14,2.79)
      \psline[linewidth=2pt]{<-}(7.24,1.11)(7.14,1.21)
      \rput[cc]{0}(7.7,3.05){$\displaystyle l$}
      \rput[cc]{0}(7.7,0.95){$\displaystyle \phi$}
  \endpspicture
\end{center}

%% file: modfig.tex
\font\thinlinefont=cmr5
\begingroup\makeatletter\ifx\SetFigFont\undefined
\def\x#1#2#3#4#5#6#7\relax{\def\x{#1#2#3#4#5#6}}%
\expandafter\x\fmtname xxxxxx\relax \def\y{splain}%
\ifx\x\y   
\gdef\SetFigFont#1#2#3{%
  \ifnum #1<17\tiny\else \ifnum #1<20\small\else
  \ifnum #1<24\normalsize\else \ifnum #1<29\large\else
  \ifnum #1<34\Large\else \ifnum #1<41\LARGE\else
     \huge\fi\fi\fi\fi\fi\fi
  \csname #3\endcsname}%
\else
\gdef\SetFigFont#1#2#3{\begingroup
  \count@#1\relax \ifnum 25<\count@\count@25\fi
  \def\x{\endgroup\@setsize\SetFigFont{#2pt}}%
  \expandafter\x
    \csname \romannumeral\the\count@ pt\expandafter\endcsname
    \csname @\romannumeral\the\count@ pt\endcsname
  \csname #3\endcsname}%
\fi
\fi\endgroup
\mbox{\beginpicture
\setcoordinatesystem units <1.00000cm,1.00000cm>
\unitlength=1.00000cm
\linethickness=1pt
\setplotsymbol ({\makebox(0,0)[l]{\tencirc\symbol{'160}}})
\setshadesymbol ({\thinlinefont .})
\setlinear
%
%
\linethickness= 0.500pt
\setplotsymbol ({\thinlinefont .})
\putrule from  5.239 25.718 to  6.826 25.718
%
%
\linethickness= 0.500pt
\setplotsymbol ({\thinlinefont .})
\putrule from  5.239 25.559 to  6.826 25.559
%
%
\linethickness= 0.500pt
\setplotsymbol ({\thinlinefont .})
\putrule from  5.239 25.400 to  6.826 25.400
%
%
\linethickness= 0.500pt
\setplotsymbol ({\thinlinefont .})
\putrule from  5.239 25.241 to  6.826 25.241
%
%
\linethickness= 0.500pt
\setplotsymbol ({\thinlinefont .})
\putrule from  5.239 25.082 to  6.826 25.082
%
%
\linethickness= 0.500pt
\setplotsymbol ({\thinlinefont .})
\putrule from  5.239 24.924 to  6.826 24.924
%
%
\linethickness= 0.500pt
\setplotsymbol ({\thinlinefont .})
\putrule from  5.239 24.765 to  6.826 24.765
%
%
\linethickness= 0.500pt
\setplotsymbol ({\thinlinefont .})
\putrule from  5.239 24.606 to  6.826 24.606
%
%
\linethickness= 0.500pt
\setplotsymbol ({\thinlinefont .})
\putrule from  5.239 24.448 to  6.826 24.448
%
%
\linethickness= 0.500pt
\setplotsymbol ({\thinlinefont .})
\putrule from  5.239 24.289 to  6.826 24.289
%
%
\linethickness= 0.500pt
\setplotsymbol ({\thinlinefont .})
\putrule from  5.239 24.130 to  6.826 24.130
%
%
\linethickness= 0.500pt
\setplotsymbol ({\thinlinefont .})
\putrule from  5.239 23.971 to  6.826 23.971
%
%
\linethickness= 0.500pt
\setplotsymbol ({\thinlinefont .})
\putrule from  5.239 23.812 to  6.826 23.812
%
%
\linethickness= 0.500pt
\setplotsymbol ({\thinlinefont .})
\putrule from  5.239 23.495 to  6.826 23.495
%
%
\linethickness= 0.500pt
\setplotsymbol ({\thinlinefont .})
\putrule from  5.239 23.336 to  6.826 23.336
%
%
\linethickness= 0.500pt
\setplotsymbol ({\thinlinefont .})
\putrule from  5.239 23.178 to  6.826 23.178
%
%
\linethickness= 0.500pt
\setplotsymbol ({\thinlinefont .})
\putrule from  5.239 23.654 to  6.826 23.654
%
%
\linethickness= 0.500pt
\setplotsymbol ({\thinlinefont .})
\putrule from  7.144 25.876 to  8.731 25.876
%
%
\linethickness= 0.500pt
\setplotsymbol ({\thinlinefont .})
\putrule from  7.144 25.718 to  8.731 25.718
%
%
\linethickness= 0.500pt
\setplotsymbol ({\thinlinefont .})
\putrule from  7.144 25.559 to  8.731 25.559
%
%
\linethickness= 0.500pt
\setplotsymbol ({\thinlinefont .})
\putrule from  7.144 25.400 to  8.731 25.400
%
%
\linethickness= 0.500pt
\setplotsymbol ({\thinlinefont .})
\putrule from  7.144 25.241 to  8.731 25.241
%
%
\linethickness= 0.500pt
\setplotsymbol ({\thinlinefont .})
\putrule from  7.144 25.082 to  8.731 25.082
%
%
\linethickness= 0.500pt
\setplotsymbol ({\thinlinefont .})
\putrule from  7.144 24.924 to  8.731 24.924
%
%
\linethickness= 0.500pt
\setplotsymbol ({\thinlinefont .})
\putrule from  7.144 24.765 to  8.731 24.765
%
%
\linethickness= 0.500pt
\setplotsymbol ({\thinlinefont .})
\putrule from  7.144 24.606 to  8.731 24.606
%
%
\linethickness= 0.500pt
\setplotsymbol ({\thinlinefont .})
\putrule from  7.144 24.448 to  8.731 24.448
%
%
\linethickness= 0.500pt
\setplotsymbol ({\thinlinefont .})
\putrule from  7.144 24.289 to  8.731 24.289
%
%
\linethickness= 0.500pt
\setplotsymbol ({\thinlinefont .})
\putrule from  7.144 24.130 to  8.731 24.130
%
%
\linethickness= 0.500pt
\setplotsymbol ({\thinlinefont .})
\putrule from  7.144 23.971 to  8.731 23.971
%
%
\linethickness= 0.500pt
\setplotsymbol ({\thinlinefont .})
\putrule from  7.144 23.812 to  8.731 23.812
%
%
\linethickness= 0.500pt
\setplotsymbol ({\thinlinefont .})
\putrule from  7.144 23.495 to  8.731 23.495
%
%
\linethickness= 0.500pt
\setplotsymbol ({\thinlinefont .})
\putrule from  7.144 23.336 to  8.731 23.336
%
%
\linethickness= 0.500pt
\setplotsymbol ({\thinlinefont .})
\putrule from  5.239 25.876 to  6.826 25.876
%
%
\linethickness= 0.500pt
\setplotsymbol ({\thinlinefont .})
\putrule from  7.144 23.178 to  8.731 23.178
%
%
\put{$\zeta(\Omega,\omega)$} [lB] at  3.969 23.947
%
%
\linethickness= 0.500pt
\setplotsymbol ({\thinlinefont .})
\putrule from  7.144 23.654 to  8.731 23.654
%
%
\linethickness= 0.500pt
\setplotsymbol ({\thinlinefont .})
\putrule from  9.049 25.876 to 10.636 25.876
%
%
\linethickness= 0.500pt
\setplotsymbol ({\thinlinefont .})
\putrule from  9.049 25.718 to 10.636 25.718
%
%
\linethickness= 0.500pt
\setplotsymbol ({\thinlinefont .})
\putrule from  9.049 25.559 to 10.636 25.559
%
%
\linethickness= 0.500pt
\setplotsymbol ({\thinlinefont .})
\putrule from  9.049 25.400 to 10.636 25.400
%
%
\linethickness= 0.500pt
\setplotsymbol ({\thinlinefont .})
\putrule from  9.049 25.241 to 10.636 25.241
%
%
\linethickness= 0.500pt
\setplotsymbol ({\thinlinefont .})
\putrule from  9.049 25.082 to 10.636 25.082
%
%
\linethickness= 0.500pt
\setplotsymbol ({\thinlinefont .})
\putrule from  9.049 24.924 to 10.636 24.924
%
%
\linethickness= 0.500pt
\setplotsymbol ({\thinlinefont .})
\putrule from  9.049 24.765 to 10.636 24.765
%
%
\linethickness= 0.500pt
\setplotsymbol ({\thinlinefont .})
\putrule from  9.049 24.606 to 10.636 24.606
%
%
\linethickness= 0.500pt
\setplotsymbol ({\thinlinefont .})
\putrule from  9.049 24.448 to 10.636 24.448
%
%
\linethickness= 0.500pt
\setplotsymbol ({\thinlinefont .})
\putrule from  9.049 24.289 to 10.636 24.289
%
%
\linethickness= 0.500pt
\setplotsymbol ({\thinlinefont .})
\putrule from  9.049 24.130 to 10.636 24.130
%
%
\linethickness= 0.500pt
\setplotsymbol ({\thinlinefont .})
\putrule from  9.049 23.971 to 10.636 23.971
%
%
\linethickness= 0.500pt
\setplotsymbol ({\thinlinefont .})
\putrule from  9.049 23.812 to 10.636 23.812
%
%
\linethickness= 0.500pt
\setplotsymbol ({\thinlinefont .})
\putrule from  9.049 23.495 to 10.636 23.495
%
%
\linethickness= 0.500pt
\setplotsymbol ({\thinlinefont .})
\putrule from  9.049 23.336 to 10.636 23.336
%
%
\linethickness= 0.500pt
\setplotsymbol ({\thinlinefont .})
\putrule from  9.049 23.178 to 10.636 23.178
%
%
\linethickness= 0.500pt
\setplotsymbol ({\thinlinefont .})
\putrule from  9.049 23.654 to 10.636 23.654
%
%
\linethickness= 0.500pt
\setplotsymbol ({\thinlinefont .})
\setdots < 0.0952cm>
\setsolid
%
%
\plot  3.913 23.812  3.651 23.812  3.882 23.689 /
\setdots < 0.0952cm>
\plot  3.651 23.812  4.921 23.495 /
\setsolid
%
%
\plot  4.659 23.495  4.921 23.495  4.690 23.618 /
\setdots < 0.0952cm>
%
%
\linethickness= 0.500pt
\setplotsymbol ({\thinlinefont .})
\setsolid
\putrule from  1.746 23.971 to  3.334 23.971
%
%
\linethickness= 0.500pt
\setplotsymbol ({\thinlinefont .})
\setdots < 0.0952cm>
\setsolid
%
%
\plot  3.850 24.300  3.651 24.130  3.907 24.187 /
\setdots < 0.0952cm>
\plot  3.651 24.130  4.921 24.765 /
\setsolid
%
%
\plot  4.722 24.595  4.921 24.765  4.666 24.708 /
\setdots < 0.0952cm>
%
%
\linethickness= 0.500pt
\setplotsymbol ({\thinlinefont .})
\setsolid
\putrule from 11.049 23.178 to 11.176 23.178
%
%
\linethickness= 0.500pt
\setplotsymbol ({\thinlinefont .})
%
%
\plot 11.176 26.204 11.113 26.458 11.049 26.204 /
\putrule from 11.113 26.458 to 11.113 22.701
%
%
\linethickness= 0.500pt
\setplotsymbol ({\thinlinefont .})
\setdots < 0.0952cm>
\plot  6.032 26.511  6.032 26.035 /
%
%
\linethickness= 0.500pt
\setplotsymbol ({\thinlinefont .})
\plot  7.938 26.511  7.938 26.035 /
%
%
\linethickness= 0.500pt
\setplotsymbol ({\thinlinefont .})
\plot  9.842 26.511  9.842 26.035 /
%
%
\put{$E$} [lB] at  1.270 23.812
%
%
\put{$\omega$} [lB] at 11.271 26.194
%
%
\put{$\Omega_1$} [lB] at  5.874 22.701
%
%
\put{$\Omega_2$} [lB] at  7.779 22.701
%
%
\put{$\Omega_3$} [lB] at  9.684 22.701
%
%
\put{$\omega_c$} [lB] at 11.271 23.114
\linethickness=0pt
\putrectangle corners at  1.270 26.537 and 11.271 22.676
\endpicture}